# Superconducting proximity effect in a Rashba-type surface state of Pb/Ge(111)

[1]Department of Physics, University of Tokyo, Bunkyo, Tokyo 113-0033, Japan
[2]Institute of Automation and Control Processes, 690041, Vladivostok, Russia

H. Huang[1*], H. Toyama[1], L. V. Bondarenko[2], A. Y. Tupchaya[2], D. V. Gruznev[2], A. Takayama[1+], R. Hobara[1], R. Akiyama[1], A. V. Zotov[2], A. A. Saranin[2], and S. Hasegawa[1]

* huanghr@surface.phys.s.u-tokyo.ac.jp
+ Present address: Waseda University, 3-4-1 Okubo Shinjuku-ku, Tokyo 169-8555, Japan.

The Rashba superconductor, in which spin-splitting bands become superconducting, is fascinating as a novel superconducting system in low-dimensional systems. Here, we present the results of *in-situ* transport measurements on a Rashba-type surface state of the striped incommensurate (SIC) phase of a Pb atomic layer on Ge(111) surface with additional Pb islands/clusters on it. We found that two-step superconducting transitions occurred at around 7 K and 3 K. The latter superconducting transition is suggested to be induced at the non-superconducting Rashba SIC area because of the lateral proximity effect caused by the superconducting Pb clusters. Our results propose a new type of Rashba superconductor, which is a platform to understand the Rashba superconducting systems.

## 1. Introduction

Researches on two-dimensional (2D) superconductors have greatly progressed in recent years thanks to various kinds of highly crystalline atomic-layer materials found [1][2] and *in situ* transport measurement techniques in ultrahigh vacuum (UHV) at low temperature [3]. There are many intriguing phenomena already found in the 2D superconducting systems, for example, the superconductor-insulator transition (SIT) in Bi thin films and others [4], the SIT mediated by a quantum metal phase [5][6], and higher critical temperatures ($T_c$) than those of bulk materials [7][8][9]. Particularly, the Rashba superconductor where spin-split bands due to the Rashba effect become superconducting, has attracted much attention as a possible unconventional 2D superconductor because the spin-triplet pairing and the Fulde–Ferrell–Larkin–Ovchinnikov (FFLO) state are theoretically predicted [10][11].

Some Rashba superconductors have been reported so far and are roughly classified into two types: one is the intrinsic superconductor in Rashba-type bands, such as (Tl, Pb) and (Au, Tl) alloy atomic layers on Si(111) [12][13], and the other is a heterojunction-type Rashba superconductor like the $LaAlO_3/SrTiO_3$ quantum well [14]. Both are induced at surfaces or interfaces of crystals where space-inversion symmetry is broken down.

The Ge(111) surface covered with 4/3 monolayer (ML) of Pb, which is called β-$\sqrt{3}\times\sqrt{3}$ structure or the striped incommensurate (SIC) phase, has a metallic band with giant-Rashba-type spin splitting [15]. The SIC phase shows no sign of superconductivity down to 0.5 K [16][17]. On the other hand, the bulk Pb is a typical superconducting material with $T_c$ = 7.2 K. By depositing the Pb nano-structures on Ge(111), we propose here the third type of Rashba superconductors, *i.e.*, a homojunction system with lateral superconducting proximity effect. The homojunction is made of the different phases of Pb; one is the non-superconducting SIC surface phase having the Rashba-type band, and the other is the bulk-like Pb islands and clusters providing Cooper pairs spilling over the SIC phase.

In this paper, we report the superconducting properties of Pb-covered Ge(111) surface, investigated by *in-situ* four-point-probe (4PP) conductivity measurements in UHV. In order to discuss the transport properties combined with the surface structure and electronic states, we also performed the scanning tunneling microscopy (STM) and the angle-resolved photoelectron spectroscopy (ARPES) in separate UHV systems. As the results of the Pb-coverage dependence of the sheet resistivity, two-step superconducting transitions were observed, where both large islands (a few hundred nm or larger in size) and small clusters (10 nm or smaller in size) existed on the SIC surface. While one of the transitions around 7 K comes from the large Pb islands, we conclude that the observed resistance drop below 3 K is induced by the superconducting current flowing through the non-superconducting Rashba-type SIC phase, which is due to the proximity effect from the superconducting Pb islands/clusters on the SIC phase.

## 2. Method

We first prepared a clean surface of Ge(111) crystal wafer of *n*-type, the resistivity of which was 40-65 Ω·cm at room temperature (RT), by several cycles of 1.0 keV Ar+ sputtering for 20 minutes and annealing at 870 K for 20 minutes in UHV, resulting in the well-known Ge(111)-c(2×8) reconstruction. Then the Pb was deposited on the Ge(111) at RT. The deposition amount was determined by the deposition duration time at a constant evaporation rate, which had been calibrated by the formation of the Si(111)-SIC-Pb phase at 4/3 ML coverage [18][19].

Figures 1(a)-(c) show the reflection high-energy electron diffraction (RHEED) patterns of Pb on Ge(111) obtained at RT for 0, 1.33 and 3.33 ML coverages of Pb, respectively. We clearly see spots from the Ge(111)-c(2×8) reconstruction in Fig. 1(a) where (11) and ($\bar{1}\bar{1}$) streaks from Ge(111) are pointed by blue arrows. In the case of 1.33 ML coverage, as shown in Fig.1(b), streaks of the $\sqrt{3}\times\sqrt{3}$ periodicity appear. Unlike the spot-like patterns of the Ge(111)-β$\sqrt{3}\times\sqrt{3}$-Pb with post annealing, this streak structure is very similar to the RHEED pattern of the Si(111)-SIC-Pb phase [20], where the (2/3 2/3) streaks (olive green arrows) are brighter than (1/3 1/3) streaks (light green arrows). This was an evidence that we acquired the Ge(111)-SIC-Pb surface structure. With additional Pb deposition up to 3.33 ML coverage, new streaks appeared outside

the (11) and $(\bar{1}\bar{1})$ streaks [red arrows in Fig. 1(c)] while the SIC streaks still remained. By measuring the ratio of distance from $(\bar{1}\bar{1})$ to (11) streaks between the red and blue arrows in the RHEED patterns (Figs. 1(a,c)), we estimated the lattice constant for the red streaks as 4.92 Å with reference of the lattice constant of Ge(111) (5.65 Å). This value is consistent with the lattice constant on the Pb(111) plane (4.92 Å), meaning that the excess Pb atoms form three-dimensional islands on the Ge(111)-SIC-Pb phase with Pb(111) face on top. The epitaxial growth relation between the Pb islands and Ge(111) substrate is Pb(111)//Ge(111), Pb$[\bar{1}10]$//Ge$[\bar{1}10]$. The reciprocal lattice of this system is shown in Fig. 1(d). The samples were prepared in the same way with the help of these RHEED patterns in the 4PP, STM, and ARPES chambers.

The STM measurements were performed at RT by the constant-current mode (Omicron MULTIPROBE system) with a homemade PtIr tip. The STM images were processed by the free software WSxM 4.0 Beta 9.1 [21].

The ARPES measurements were carried out at 100 K with a hemispherical electron energy analyzer (MBS-A1) using a photon energy *hv* = 14 eV at the beamline BL13 at Saga Light Source [22]. We set the energy and momentum resolution at 20 meV and 0.015 Å$^{-1}$, respectively.

The transport measurements were performed *in situ* in UHV by the micro-4PP measurement system (UNISOKU USM-1300S) [23], in which the sample was cooled down to 0.85 K under the magnetic field up to 7 T applied perpendicularly to the sample surface. We used a homemade 4PP which consisted of four copper wires of 100 μm in diameter, aligning on a line with the probe spacing of ~200 μm. Since we made soft but direct contact between the probes and the sample surface, the contact areas are on the order of micrometers in size at least, which is much larger than the diameter of the Pb clusters/islands and the distance between them as mentioned later. Therefore, our transport measurement system does not make possible to measure separately the resistance on the Pb clusters/islands and that of the SIC-Pb area between them. The resistivity here is always an average value of the area having a few hundred μm in size.

All measurements of STM, ARPES, and 4PP were done *in situ* for samples prepared in the UHV chambers without exposing them to air; the sample preparation and the measurements were done in the same UHV chambers. Therefore. we do not need a capping layer on the sample surface to protect the surface from oxidation for the transport measurements, which is an advantage of our method over *ex situ* measurements.

## 3. Results and discussion

### 3. 1. Atomic structures: STM

First, we confirmed the c(2×8) reconstruction of the clean surface of Ge(111) in a large area by STM. The terraces of the Ge(111), together with the atomic resolution of the reconstruction can be seen from the STM images in Fig 1(a). Then Pb was deposited on it. As shown in Fig. 2(b), we see clearly the $\sqrt{3}\times\sqrt{3}$ structure on the terrace for 1.33

ML-Pb coverage as in the previous studies [24] and the RHEED pattern in Fig. 1(b). As better illustrated with enhanced contrast in Fig. 2(b) (lower), the stripe patterns of the SIC phase can be seen. The distance between two adjacent stripes is around 12 nm, which is quite similar to the previous report of the Ge(111)-SIC-Pb [24]. Therefore, it is confirmed that our sample is the SIC phase with 1.33 ML Pb deposition. In Fig. 2(c), we find that with a little extra amount of Pb deposition on the Ge(111)-SIC-Pb, two kinds of small islands are formed on the terrace: one is relatively large islands (ca. 10 nm in size) with the $\sqrt{3}\times\sqrt{3}$ periodicity on it, and the other is small Pb clusters (a few nm in size) with a characteristic pattern on it. The former is Pb-covered Ge(111) islands of one-unit-layer higher due to the mismatch in the number density of Ge atoms between the c(2×8) and SIC structures, while the latter is Pb clusters. The islands with the $\sqrt{3}\times\sqrt{3}$ periodicity on it already appeared at 1.33 ML coverage, meaning that this is not due to the extra Pb atoms over 1.33 ML.

With further deposition of Pb up to 3.33 ML, as shown in Fig. 2(d), large Pb(111) islands (a few hundred nm or more in size) having the Pb(111)-1×1 structure on the top appear. This is consistent with the RHEED pattern in Fig. 1(c). The height of these islands varies from 5.5 nm to 6.3 nm. The average height of these islands is ~6 nm, corresponding to ~17 atomic layers of Pb. Figure 2 (e) is an enlarged view of the green square in Fig. 2 (d). It should be noted that the $\sqrt{3}\times\sqrt{3}$ Pb islands and the Pb clusters as in Fig. 2(c) still remain on the ravines between the big Pb(111) islands, and the density of the small Pb islands/clusters does not change even with further Pb deposition; the big Pb(111) islands just grow in size.

## 3. 2. Electronic structures: ARPES

Next, we investigated the electronic states for the 1.3 and 3.0 ML-Pb/Ge(111) samples. Figure 3(a) shows the experimentally determined Fermi surface of the Ge(111)-SIC-Pb phase (1.3 ML). This looks similar to that at the previous studies for Ge(111)-$\sqrt{3}\times\sqrt{3}$-Pb which is formed by post annealing after Pb deposition [15][16]; one can see large hexagonal Fermi surfaces (FS) near the zone boundary centered at the $\bar{\Gamma}$ point of the first and second surface Brillouin zones (SBZ). It should be noted that the FS is more clearly observed at the second SBZ for the sake of the matrix element.

Figures 3(b) and (c) show the band dispersion along the $\bar{\Gamma}$ $\bar{\mathrm{M}}$ $\bar{\Gamma}$ line which was obtained by taking the second derivative of the energy distribution curve (EDC) and the momentum distribution curve (MDC), respectively. These clearly show the band splitting near the Fermi level $E_F$ around $\bar{\mathrm{M}}$ point. The band dispersion here is very similar to those for Ge(111)-$\sqrt{3}\times\sqrt{3}$-Pb phase [15]. This is natural because the SIC-Pb phase is essentially composed of the small $\sqrt{3}\times\sqrt{3}$-Pb domains separated by incommensurate domain walls as seen in Fig. 2(b)(lower) [24]. On the other hand, the Fermi surface (Fig. 3(a)), which is the intensity mapping of photoelectrons, not the second derivative, does not look splitting, as opposed to that in Ref. [15]. This may be partially because of the poorer resolution of our measurements, and also because of the smaller domain size of $\sqrt{3}\times\sqrt{3}$-Pb domains in the SIC-Pb phase which makes the bands blurred compared with those in Ref. [15]. The band splitting shown in Figs. 3(b,c) clearly indicate the Rashba effect occurs at Ge(111)-SIC-Pb phase as at Ge(111)-√3×√3-Pb phase in Ref. [15] though the Fermi surface mappings in Fig. 3(a) does not look splitting. Judging from the band dispersion in Figs. 3(b, c) and previous analysis in Ref. [15], it is conceivable that the hexagonal FS in Fig. 3(a) is composed of two concentric hexagons with opposite spin directions.

In the case of 3 ML Pb coverage, we observed not only the hexagonal FS at the 2nd SBZ, but also a circular FS centered at the $\bar{\Gamma}$ point as shown in Fig. 3(d); the circular FS is so large that it is just outside of the 1st SBZ and very close to the hexagonal FSs in the second SBZ. The band corresponding to the circular FS almost overlaps the Rashba bands from around $\bar{M}$ point (Figs. 3(e, f)). This circular FS and corresponding band are attributed to the electronic state from the Pb islands [25]. At first sight, it is difficult to discuss whether the Rashba splitting bands still survive only from the ARPES data presented here, because the Fermi velocity and Fermi wave vector of both bands are very similar. Judging from our STM measurement in Fig. 2(d) in which the SIC-Pb phase is still exposed at the areas between the Pb islands/clusters, however, it is suggested that the Rashba splitting bands survive and coexist with the band from the Pb islands.

### 3. 3. Electronic transport: 4PP

**<Weak anti-localization>**

Then we measured the temperature dependence of sheet resistance $R_{\mathrm{sheet}}$ for the Ge(111)-SIC-Pb phase (1.33 ML) below 6 K, as shown in Fig. 4(a), where the conductivity of the Ge substrate can be neglected due to freeze-out of carriers there. The resistivity gradually increases with the temperature decreasing. Below around 1.5 K, the increasing rate becomes higher. There is no sign of superconductivity down to 0.85 K.

Then, the magnetoresistance was measured at 1 K as a function of the applied magnetic field up to 7 T. Figure 4(b) shows the result by converting the resistivity $R_{\mathrm{sheet}}$ to sheet conductivity $\sigma_{\mathrm{sheet}}$. As the magnetic field increases, $\sigma_{\mathrm{sheet}}$ decreases. However, in the low-magnetic-field range (0~0.5 T), the field dependence of the conductance is different from that in the high-magnetic-field range: the curve shape of the conductivity versus magnetic field in the low-magnetic-field range is convex as shown in Fig. 4(c), while the shape in the high-magnetic-field range is concave. So, these two areas show different types of magnetoresistance; the latter is classical magnetoresistance effect while the former is weak anti-localization (WAL) effect as described below.

Here, we mainly focus on the low-magnetic-field range, where the relation between $\sigma_{\mathrm{sheet}}$ and the magnetic field can be well fitted by the Hikami-Larkin-Nagaoka equation as below [26]:

$$\Delta\sigma = \sigma(B) - \sigma(0) = \frac{\alpha\, e^2}{2\,\pi^2\,\hbar}\left[\psi\left(\frac{1}{2} + \frac{\hbar}{4\,e\,l_\phi^2\,B}\right) - \ln\left(\frac{\hbar}{4\,e\,l_\phi^2\,B}\right)\right], \qquad (1)$$

where $\alpha$, $e$, $\hbar$, $B$ and $l_\phi$ are the number of conduction channels, the elementary electric charge, the reduced Planck's constant, the perpendicular magnetic field applied, and the phase coherence length, respectively. The fitting parameters were $l_\phi$ and $\alpha$. We estimated the value of the $l_\phi$ and $\alpha$ as 211 nm and -0.48 by fitting Eq. (1) as shown by a dotted curve in Fig. 4(c). The fitting result is in good agreement with the expectation of the WAL. The value of α is very close to the theoretical value -0.5 for a single-channel WAL [26]. The WAL behavior comes from the strong SOC of the Ge(111)-SIC-Pb phase, which is confirmed also by the ARPES measurements as the Rashba effect as described in Fig. 3 and Ref. [15].

**<Superconductivity>**

We also performed the transport measurement on the sample with 3.3 ML-Pb deposition which was Ge(111)-SIC-Pb phase coexisting with Pb islands/clusters on it as shown in Figs. 2(d,e). Figure 5(a) shows the $R_{\mathrm{sheet}}$ as a function of temperature under some different magnetic fields between 0 and 1 T. From the data in the zero magnetic field, we observed that $R_{\mathrm{sheet}}$ drops steeply at two temperatures, around 7 K and below 3 K. These two drops become broader and shift to the lower temperature as the magnetic field increases. The drops at 7 K and 3 K vanish at 0.4 and 0.1 T, respectively. The results of the magnetoresistance measurements at different temperatures are summarized in Fig. 5(b).

Because the results seem consistent with a picture that the superconductivity is broken by applying magnetic field to return to the normal state having a finite resistance, it is reasonable to assume that these drops are superconducting transitions with different $T_{\mathrm{c}}$ and different critical magnetic fields ($H_{\mathrm{c}}$) although the resistivity does not reach zero at the lowest temperature 0.85 K we can reach.

Here, we first focus on the resistance drop around 7 K. We define $T_{\mathrm{c}}$ as the temperature at which $R_{\mathrm{sheet}}$ becomes 90 % of the normal resistance [27]. From the thermodynamics, the relationship between $H_{\mathrm{c}}$ and temperature obeys the Tuyn's law [28]:

$$H_c(T) = H_c(0)\left[1 - \left(\frac{T}{T_c}\right)^2\right], \qquad (2)$$

where $H_{\mathrm{c}}(0)$ is $H_{\mathrm{c}}$ at zero temperature. The relation between $H_{\mathrm{c}}$ and the temperature was depicted in Fig. 5(c). As the result of fitting by Eq. (2), the intercept of the temperature axis was estimated to be 7.1 K, which is almost the same as the $T_{\mathrm{c}}$ in the bulk Pb (7.2 K) [28]. This means that the resistance drop around 7 K is due to superconducting transition of the large Pb(111) islands formed on the SIC phase as shown in Fig. 2(d). Since the Pb(111) islands are not connected each other, the resistance does not reach zero even below 7 K.

Figure 5(d) shows the temperature dependence of $R_{\mathrm{sheet}}$ at Pb coverage of 2.5, 3.3, 3.9, and 6.5 ML, respectively, without magnetic field applied. As the deposition amount increases, $R_{\mathrm{sheet}}$ in the normal state becomes lower. This is reasonable because the areal fraction occupied by the Pb(111) islands on the surface increases. Since, however, the size, height, and number density of the Pb(111) islands are not controllable, the decrease of the normal-state resistance is not simply proportional to the deposition amount. The important thing here is that the temperature dependences of $R_{\mathrm{sheet}}$ look very similar to each other for all samples with different coverages, *i.e.*, there are two drops at 7 K and below 3 K. As mentioned above, the $R_{\mathrm{sheet}}$ drop around 7 K is caused by the superconducting transition at the bulk-like Pb(111) islands. On the other hands, for the drop below 3 K, $R_{\mathrm{sheet}}$ gradually decreases above ~2 K, and steeply decreases below ~2 K with decreasing temperature. This behavior is common for 2D superconductors due to the large superconducting fluctuation [1]. By using the theoretical fitting to the Aslamazov-Larkin-Maki-Thompson correction including the effect of the 2D superconducting fluctuation as below [29], $T_c$ is given for each sample:

$$\rho = \frac{1}{\sigma_0 + \sigma_{AL} + \sigma_{MT}}, \qquad (3)$$

$$\sigma_{AL} = \frac{e^2}{16\hbar} \cdot \frac{T_c}{T - T_c}, \qquad (3a)$$

$$\sigma_{MT} = \frac{e^2}{8\hbar} \cdot \frac{T_c}{T - (1+\delta)\, T_c} \ln \frac{T - T_c}{\delta\, T_c}, \qquad (3b)$$

where $\sigma_0$ is the normal-state conductivity, $\sigma_{AL}$ is the Aslamazov-Larkin term, $\sigma_{MT}$ is the Maki-Thompson term, and $\delta$ is the pair-breaking parameter. The solid (black) curves in Figs. 5(a) and (d) are the fitted ones by Eq. (3) which agree well with the experimental data. Moreover, we found that the $T_c$ estimated by fitting by Eq. (3) is almost the same value (~0.79 K) for all the samples. Since this $T_c$ is below the lowest temperature (0.85 K) we can reach in our experimental machine, the resistance does not reach zero.

Finally, let us discuss the origin of the superconducting transition with $T_c$=0.79 K. In the previous STS study of the Ge(111)-β√3×√3-Pb surface with large Pb(111) islands on it, no superconducting gap is observed down to 0.5 K on the β√3×√3-Pb areas [17]. This means the Ge(111)-SIC-Pb phase, which is composed of the same but small √3×√3-Pb domains as Ge(111)-β√3×√3-Pb surface, is not superconducting above 0.85 K. This is also shown by our own measurement of Fig. 4(a).

The paper [17] also reports that the Cooper-pair wavefunction spills out from the superconducting Pb(111) islands onto the β√3×√3-Pb area with the maximum penetration length of ~80 nm at 0.5 K [17]. As mentioned in Fig. 2(d), on the other hand, the mean distance among the large Pb(111) islands observed by our STM measurement is over ~180 nm. Then, the Cooper pairs spilling from Pb(111) islands cannot overlap each other on our sample even if the Pb(111) islands become superconducting. Since, therefore, the superconducting area are not connected each other on the surface even with spilling out of the Cooper-pair wavefunction, this results in finite values of $R_{\mathrm{sheet}}$ even below 7 K.

Therefore, the superconducting transition with $T_c$=0.79 K cannot explained just by considering the superconducting transition of the large Pb(111) islands; it is suggested that the transition relates to the small Pb clusters observed on the SIC-Pb phase between the Pb(111) islands (Figs. 2(d) and (e)). When the temperature decreases below 3 K, the small Pb clusters on Ge(111)-SIC-Pb phase become superconducting. Generally, the $T_c$ of smaller islands is lower than that of the bulk owing to the confinement effect [28], and thus we should consider it as a nanometer-size superconductor with $T_c$ lower than that of the bulk Pb. For different amounts of Pb deposition, our STM results show that the number density of the Pb clusters on the Ge(111)-SIC-Pb phase does not change much although the size and height of the large Pb(111) islands increases with the deposition amount of Pb. This is considered as the Stranski-Krastanov growth mode, which is different from the Frank–van der Merwe growth mode of Pb on Ge(111) for the low temperature deposition of Pb [30]. The small Pb clusters are all separated from each other and the mean distance among them is ~6 nm as shown in Fig. 2(e). Hence, it can be assumed that the origin of the superconducting transition at $T_c = 0.79$ K is as follows: when the temperature decreases below 3 K, the Pb clusters become superconducting so that the surface resistance decreases. At the same time, the Cooper pairs spill out and penetrate into the Ge(111)-SIC-Pb phase. When the temperature decreases low enough, the penetration length of the Cooper pairs over the SIC-Pb phase increases and finally around 0.79 K, the penetration areas connect with each other and the whole surface becomes superconducting. Since as mentioned early, the number density of the Pb clusters on the SIC-Pb phase does not change much with increasing the Pb deposition amount, the $T_c$ has a constant value of 0.79 K irrespective of the Pb amount. To confirm this scenario, we need to observe the superconducting transition of the Pb clusters by STS, which is the subject of the future work.

**Summary**

We studied the transport properties of Pb on Ge(111) substrate with different deposition amounts, and analyzed the data with the aid of STM and ARPES results. The different transport behaviors were observed depending on the deposition amounts of Pb. With 1.33 ML deposition, which is just for formation of the striped incommensurate (SIC) phase, it is found that the sheet resistance $R_{sheet}$ increases with decreasing the temperature below 2 K, without sign of superconductivity down to 0.85 K. This is due to weak anti-localization (WAL) effect, as verified by the magnetoresistance measurements combined with fitting to the Hikami-Larkin-Nagaoka equation. With over 1.33 ML deposition of Pb, the Pb(111) islands and clusters form sparsely on the SIC-Pb surface phase. In these systems, two resistance drops were observed; one was a sudden drop around 7 K because of the superconducting transition of the large Pb(111) islands, and the other was a gradual drop below 3 K, due to the proximity-induced superconducting transition of the SIC phase with $T_c = 0.79$ K, originating from the small Pb clusters. The gradual decrease in $R_{sheet}$ below 3K was well fitted by the 2D superconducting formula including large superconducting fluctuation. The interesting finding here is that the SIC phase is not superconducting by itself, but the phase wholly becomes superconducting with help of superconducting Pb clusters on it.

**Acknowledgment**

The photoemission experiments were performed at Saga University Beamline (SAGA-LS/BL13) in Saga Light Source with a proposal of H27-305V under the support of the Ministry of Education, Culture, Sports, Science and Technology (MEXT), Japan. This work was supported by the JSPS through KAKENHI Nos. 16H02108, Bilateral Joint Research Project of JSPS and RFBR Grant No. 18-52-50022 and Russia President grant MK-343.2019.2 for young researchers.

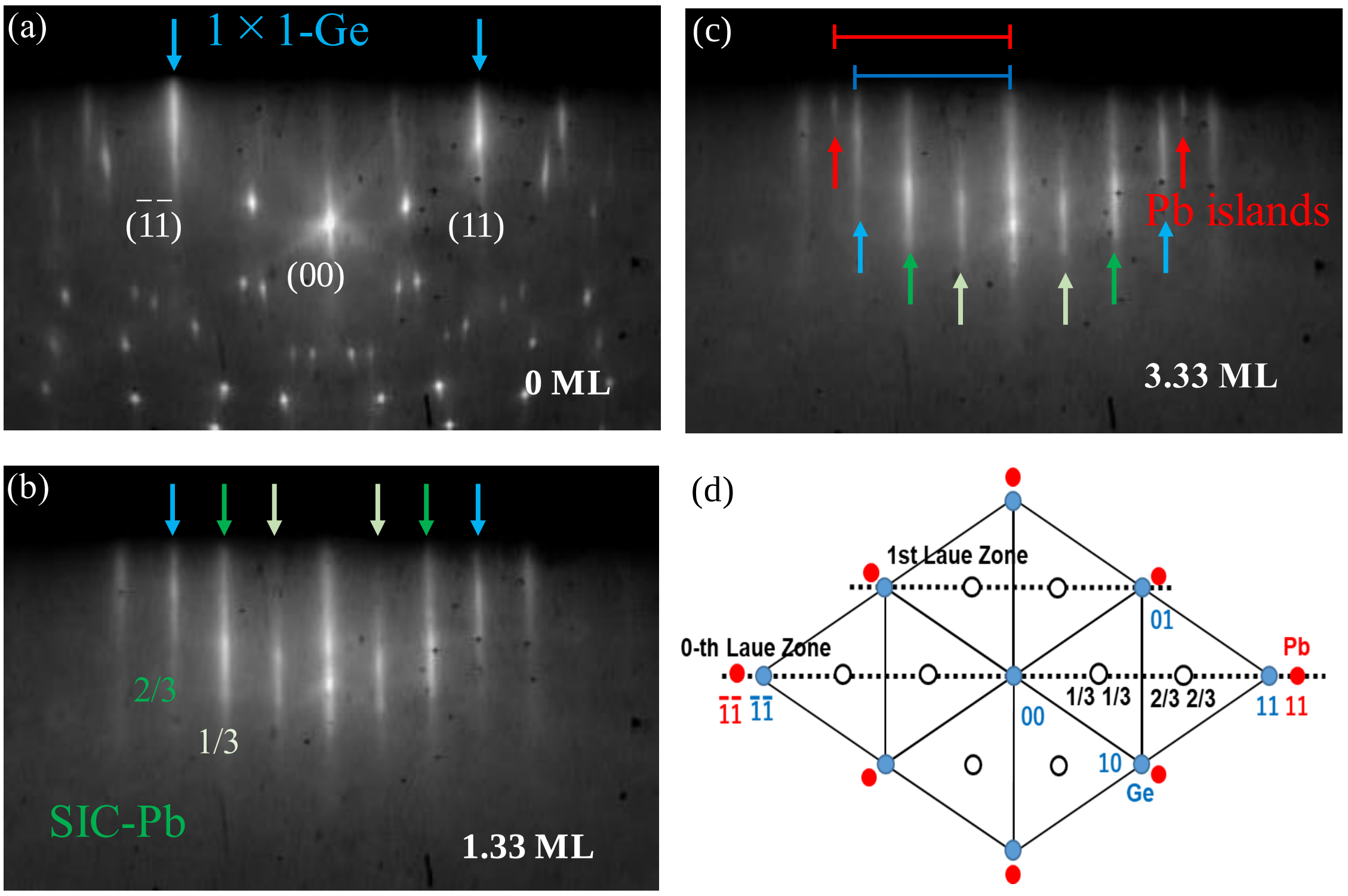

Fig 1 (a)-(c) RHEED patterns of the samples with 0, 1.33 and 3.33 ML amounts of Pb deposition, respectively, on Ge(111) substrate. (a) Ge(111)-c(2×8) clean surface reconstruction of the saubstrate. The blue arrows represent for the (11) and $(1\bar{1})$ streaks. (b) Ge(111)-SIC-Pb phase (1.33 ML Pb). The (2/3 2/3) streaks (green arrows) are stronger than the (1/3 1/3) streaks (light green arrows). (c) Ge(111)-SIC-Pb phase with additional Pb islands on it (3.33 ML Pb). Streaks indicated by red arrows come from Pb(111) islands while blue ones are from Ge(111). (d) The reciprocal lattice corresponding to (c).

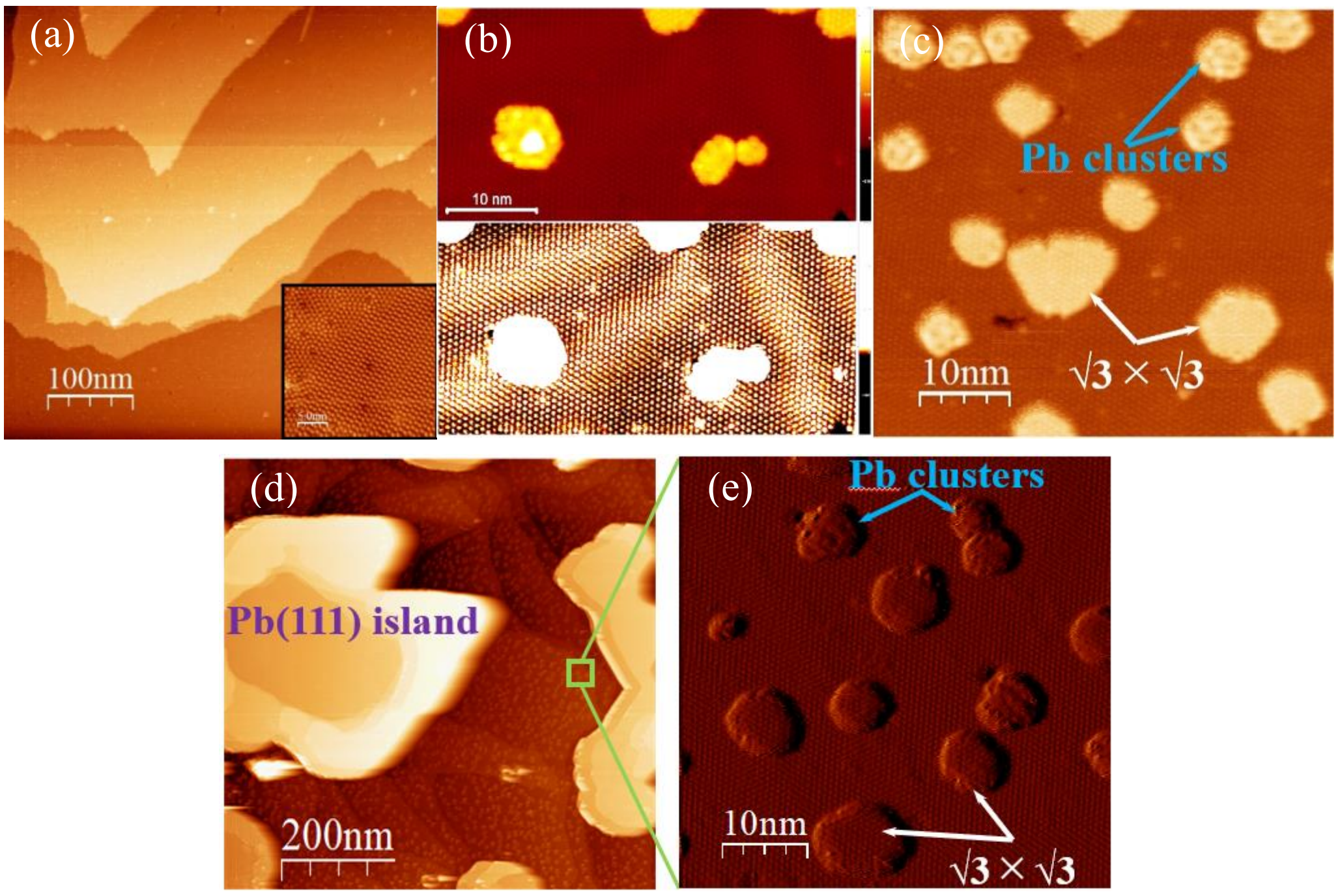

Fig 2 (a) The STM image of Ge(111)-c(2x8) clean surface of the substrate. Several terraces separated by atomic steps are seen on the substrate. The inset shows the atomic resolution of the c(2x8) reconstruction. (b) The Ge(111)-SIC-Pb phase (upper) and the contract-enhanced one (lower) of the same area. The stripes in the lower figure are incommensurate domain boundaries, characteristic of the SIC-Pb phase. (c) High-resolution STM image of the Ge(111)-SIC-Pb phase with a little extra amount of Pb deposition. Two kinds of small islands and clusters can be seen: one is with the $\sqrt{3}\times\sqrt{3}$ periodicity on it (white arrows) and the other is with a characteristic pattern on it (blue arrows). (d) Image of 3.33 ML deposition of Pb. Large islands of Pb(111) are formed on the SIC-Pb phase. (e) A derivative image of the area between the large Pb(111) islands (indicated by a square in (d)). The $\sqrt{3}\times\sqrt{3}$ islands and the Pb clusters can be seen on the SIC-Pb phase as in (c).

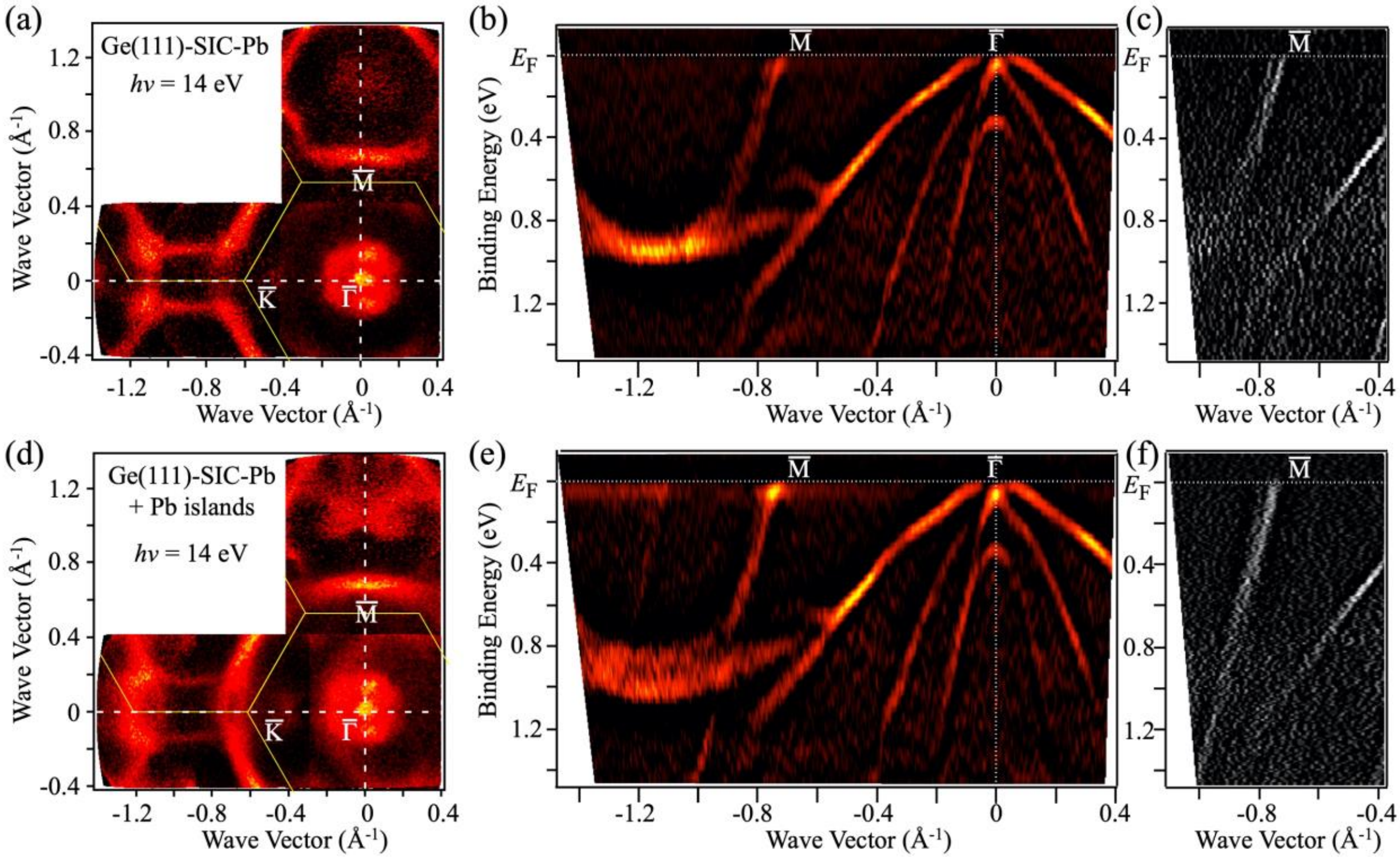

Fig 3 The ARPES results measured with $h\nu$= 14 eV. (a) The Fermi surface of the Ge(111)-SIC-Pb phase (1.3 ML Pb). The yellow solid lines show the 1×1 surface Brillouin zone of Ge(111). (b, c) Its band dispersions along the $\bar{\Gamma}$ $\bar{M}$ $\bar{\Gamma}$ line, which were obtained by taking the second derivative of the EDC and the MDC, respectively. Bright areas correspond to the bands. (d) The Fermi surface of the Ge(111)-SIC-Pb sample with Pb islands/clusters on it (3 ML Pb). (e, f) Its band dispersions along the $\bar{\Gamma}$ $\bar{M}$ $\bar{\Gamma}$ line, which were obtained by taking the second derivative of the EDC and the MDC, respectively.

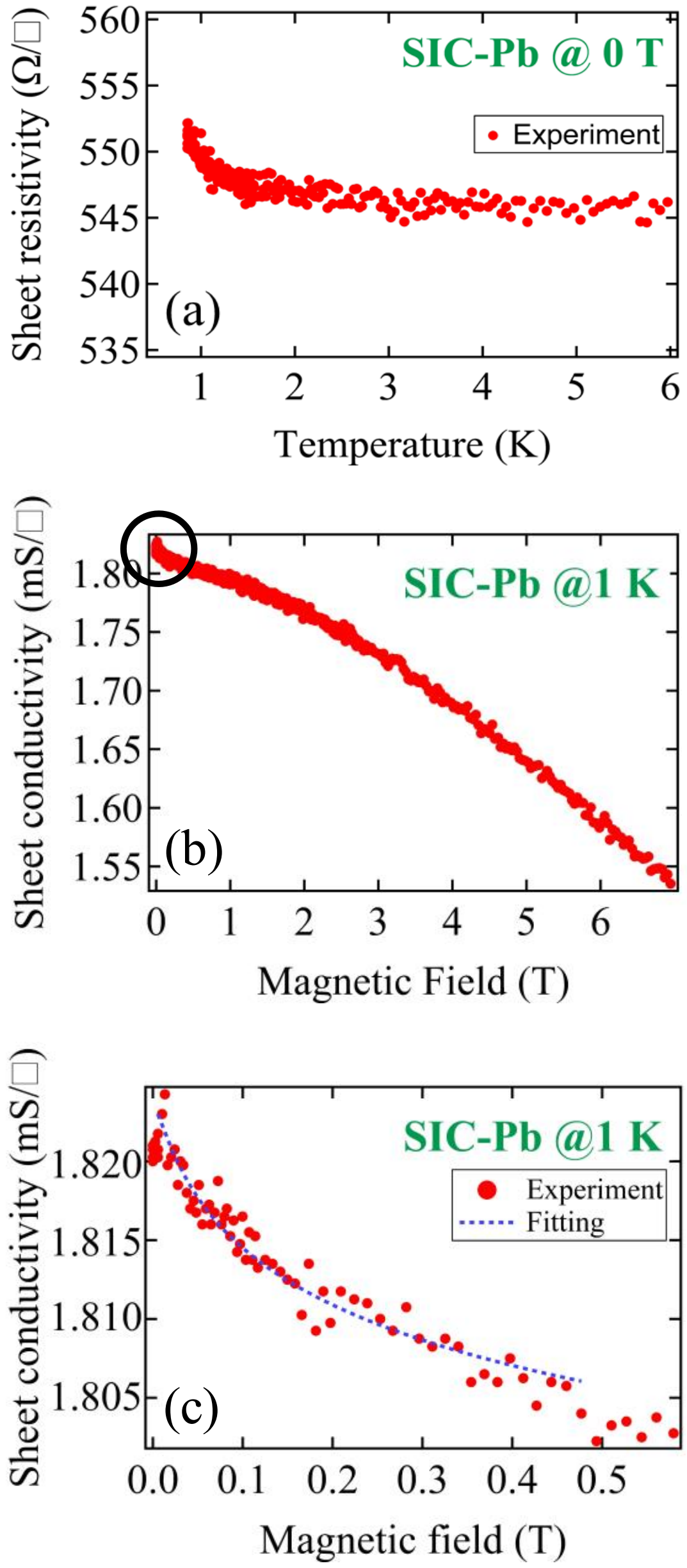

Fig 4 Transport properties of the Ge(111)-SIC-Pb phase (1.33 ML Pb). (a) Resistance-vs-temperature curve under zero magnetic field. (b) Magnetic-field dependence of the conductance at 1 K from 0 T to 7 T. (c) The enlarged figure at the range from 0 T to 0.6 T. The experimental result can be well fitted by the Hikami-Larkin-Nagaoka equation Eq. (1) (dashed curve), which indicates the WAL property.

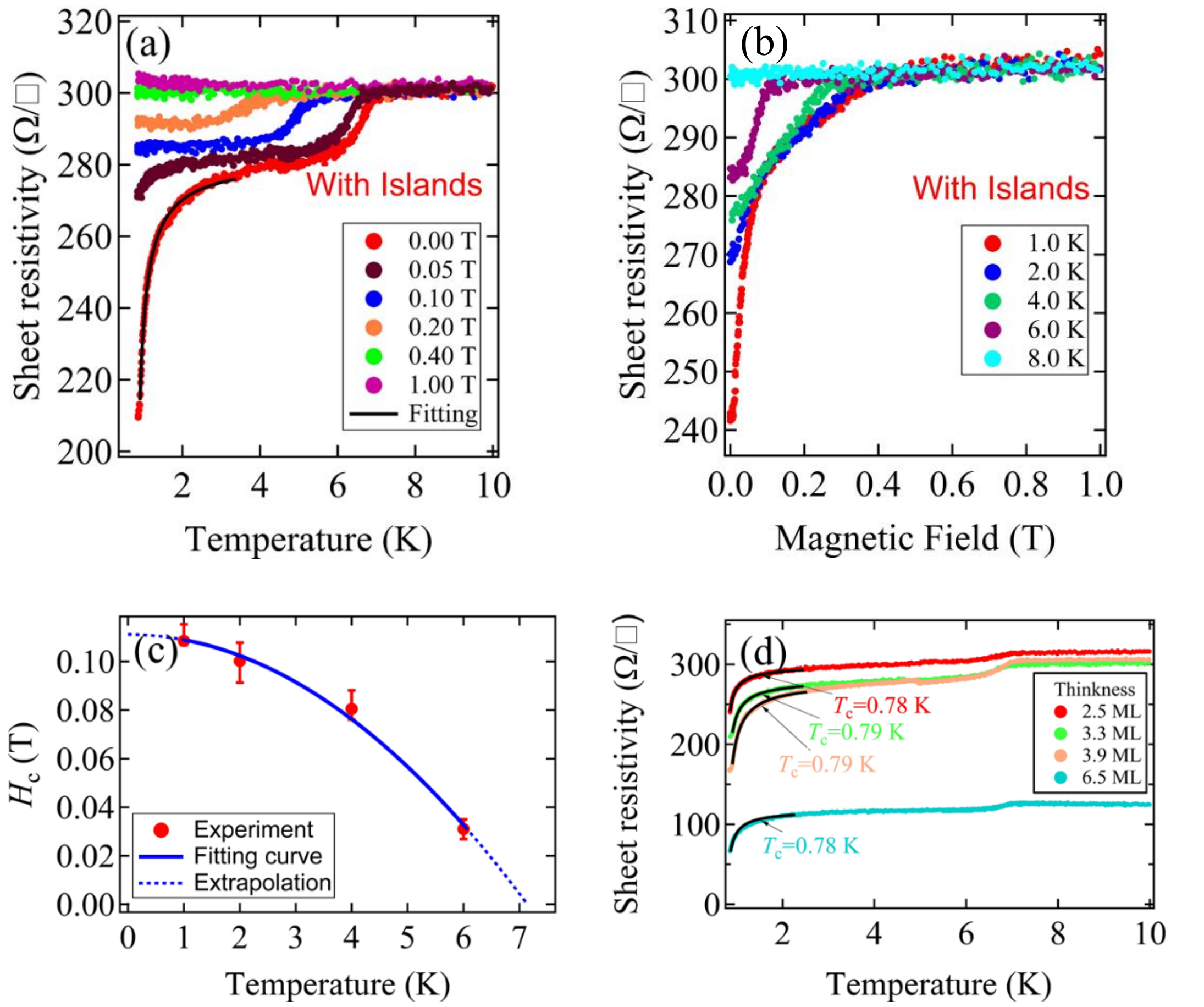

Fig 5 Superconducting behavior of the Ge(111)-SIC-Pb phase coexisting with Pb islands/clusters on it (3.3 ML Pb). (a) Temperature dependences of the sheet resistance under different magnetic fields. The curve under zero magnetic field was fitted by the 2D superconducting formula Eq. (3) (black solid curve). (b) Magnetic-field dependence of the conductance at different temperatures. With the increase of the magnetic field, the resistivity increases and finally reaches a specific value (around 300 Ω/☐), the normal-state resistivity. (c) The relationship between the temperature and the critical magnetic field. The critical magnetic field is defined as the field at which the resistance reaches the 90 % of the normal-state resistance. The experimental results are well fitted by Eq. (2). The $T_c$ can be determined by extrapolating the curve. (d) Temperature dependences of the sheet resistivity for 2.5, 3.3, 3.9 and 6.5 ML Pb coverages, respectively. All the curves were fitted by Eq. (3) (lack solid curve).